\documentclass[conference]{IEEEtran}
\IEEEoverridecommandlockouts

\usepackage{cite}
\usepackage{amsmath,amssymb,amsfonts}
\usepackage{graphicx}
\usepackage{textcomp}
\usepackage{xcolor}
\usepackage{ifthen}
\usepackage{subcaption}
\usepackage{comment}
\usepackage{algorithm}
\usepackage{algpseudocode}
\usepackage{url}
\newboolean{longversion}
\setboolean{longversion}{false} 

\usepackage{enumitem}
\def\BibTeX{{\rm B\kern-.05em{\sc i\kern-.025em b}\kern-.08em
    T\kern-.1667em\lower.7ex\hbox{E}\kern-.125emX}}
\usepackage{multirow} 
\renewcommand{\baselinestretch}{0.85}
\begin{document}

\title{REAL: Reinforcement Learning-Enabled xApps for  Experimental Closed-Loop Optimization in O-RAN with OSC RIC and srsRAN
\thanks{This material is based upon work supported by the National Science Foundation under Grant Numbers  CNS-2318726, and CNS-2232048.}
}

\author{
	\IEEEauthorblockN{
	Ryan Barker, 
        Alireza Ebrahimi Dorcheh,
        Tolunay Seyfi,
        Fatemeh Afghah}

    \IEEEauthorblockA{Holcombe Department of Electrical and Computer Engineering, Clemson University, Clemson, SC, USA \\
        Emails: \{rcbarke, alireze, tseyfi, fafghah\}@clemson.edu}

}

\maketitle


\begin{abstract}
Open Radio Access Network (O-RAN) offers an open, programmable architecture for next-generation wireless networks, enabling advanced control through AI-based applications on the near-Real-Time RAN Intelligent Controller (near-RT RIC). However, fully integrated, real-time demonstrations of closed-loop optimization in O-RAN remain scarce. In this paper, we present a complete framework that combines the O-RAN Software Community RIC (OSC RIC) with srsRAN for near-real-time network slicing using Reinforcement Learning (RL). Our system orchestrates resources across diverse slice types (eMBB, URLLC, mMTC) for up to 12 UEs. 
We incorporate GNU Radio blocks for channel modeling, including Free-Space Path Loss (FSPL), single-tap multipath, AWGN, and Doppler effects, to emulate an urban mobility scenario. Experimental results show that our RL-based xApps dynamically adapt resource allocation and maintain QoS under varying traffic demands, highlighting both the feasibility and challenges of end-to-end AI-driven optimization in a lightweight O-RAN testbed. Our findings establish a baseline for real-time RL-based slicing in a disaggregated 5G framework and underscore the need for further enhancements to support fully simulated PHY digital twins without reliance on commercial software.
\end{abstract}

\begin{IEEEkeywords}
O-RAN, 
srsRAN, 
Reinforcement Learning, Network Slicing, Resource Allocation 
\end{IEEEkeywords}

\section{Introduction and Related Work}
\label{sec:intro}

Fifth-generation (5G) cellular networks introduce a range of capabilities, including enhanced Mobile Broadband (eMBB), Ultra-Reliable Low-Latency Communications (URLLC), and massive Machine-Type Communications (mMTC) \cite{AIin5G}. These services pose diverse Quality of Service (QoS) requirements, demanding a more flexible Radio Access Network (RAN) that can dynamically allocate resources in near real-time. The Open Radio Access Network (O-RAN) framework meets this need by disaggregating traditional, vendor-locked architectures into interoperable components. Central to this vision is the near-Real-Time RAN Intelligent Controller (near-RT RIC), an AI-driven platform that executes fine-grained optimization of RAN functions over the E2 interface \cite{ORAN, ORAN_Tommaso}.
A distinguishing feature of O-RAN is the ability to host third-party applications, known as xApps, on the near-RT RIC. These xApps can employ machine learning or control-theoretic strategies to orchestrate diverse slice types, ranging from high-throughput eMBB to latency-sensitive URLLC \cite{AIinORAN}. Among learning-based approaches, \emph{Reinforcement Learning} (RL) is well-suited for adaptive, real-time control: it can iteratively refine its policy based on a reward signal reflecting slice performance (e.g., throughput, latency, or reliability). 

In recent years, several platforms have explored the integration of RL and AI in O-RAN, leveraging testbeds to address resource allocation and network optimization challenges. X5G introduces a modular 5G O-RAN platform with NVIDIA ARC for GPU-accelerated PHY layer tasks and OpenAirInterface (OAI) for higher layers \cite{villa2023x5g}. It supports real-time control via the near-RT RIC, evaluating performance using up to eight concurrent Commercial Off-the-Shelf (COTS) UEs. ORANSlice, on the other hand, emphasizes multi-slice resource allocation with compliance to 3GPP and O-RAN standards, leveraging xApps for dynamic physical resource block (PRB) allocation and scheduling policies \cite{cheng2024oranslice}. ColO-RAN offers a large-scale framework for ML-based xApp development, integrating RL-based control strategies and large-scale data collection using the Colosseum wireless network emulator \cite{coloran}. 
The PandORA framework introduces an automated approach to designing and evaluating DRL-based xApps for Open RAN, leveraging the Colosseum wireless network emulator for large-scale testing under diverse traffic and channel conditions \cite{tsampazi2024pandora}. 


\section{RELATED WORK}

Various simulation platforms and frameworks have been utilized in recent works to evaluate the effectiveness of AI/ML-based optimization strategies within O-RAN architectures. These platforms, ranging from OpenRAN Gym to NS-3 and custom simulators, offer unique features tailored to specific research objectives such as resource allocation, interference management, and network slicing. Each study highlights distinct implementation strategies, performance metrics, and scenarios.
\cite{tsampazi2023} presents an O-RAN architecture using Deep RL (DRL)-based xApps for optimizing eMBB, URLLC, and mMTC slices. Tested on the Colosseum wireless network emulator via the OpenRAN Gym framework, these xApps apply Proximal Policy Optimization (PPO) for real-time resource management. The study evaluates multiple configurations, revealing trade-offs in action spaces and reward functions, showcasing scalability and adaptive performance across varied network conditions.
Building upon this, \cite{marojevic2022actor} investigates PPO for real-time resource management within O-RAN but adds a comparative analysis of Advantage Actor-Critic (A2C). Both models optimize resource allocation through the E2 interface in communication with the near-real-time RIC, but the study demonstrates PPO's faster convergence and superior rewards over A2C. Tested in a simulated RAN environment under dynamic conditions, the research highlights PPO's superior efficiency for resource management.
In a shift toward interference management, \cite{Anand2023xApp} introduces a machine learning-based xApp for mitigating co-tier interference in a Heterogeneous Network (HetNet) environment, with a focus on improving QoE for services like video and VoIP. The xApp employs a Multi-Classification and Offloading Scheme (MLMCOS), utilizing models such as Random Forest and CNN to classify users based on interference levels and offload them to femtocells. Tested using NS-3 \cite{ns3}, this study emphasizes interference management within HetNet environments, offering a more focused solution for dense networks compared to the previous studies on resource allocation.
Further expanding on AI-driven optimization, \cite{Colosseum2023} highlights Colosseum's role as an AI/ML-based digital twin platform for O-RAN development. Through its OpenRAN Gym framework, the platform supports real-time deployment and testing of algorithms such as network slicing, scheduling, and spectrum sharing. Integrated with SDRs and real-world RF conditions, Colosseum allows for scalable, high-fidelity testing of AI/ML xApps, enabling continuous interaction between the digital twin and physical network layers. This infrastructure bridges the gap between simulation and real-world deployment, offering detailed insights into the scalability and robustness of AI/ML solutions under complex conditions.


These efforts reveal two primary gaps including 
i) \textbf{Full-stack integration at scale:}  Existing systems often handle only a small number of UEs or operate within constrained simulation setups (\cite{marojevic2022actor}, \cite{Colosseum2023}, \cite{Anand2023xApp}), and ii)
 \textbf{Real-time RL-based slicing:} Many frameworks rely on offline training or partial control loops, deferring real-time decision-making and adaptability (\cite{tsampazi2024pandora}, \cite{tsampazi2023},\cite{coloran}). Unlike many frameworks that rely on offline training, which cannot capture real-time environmental feedback, our work emphasizes online training. In this approach, the RL agent interacts directly with the srsRAN simulator during training, receiving immediate rewards based on its resource allocation decisions. This ensures dynamic adaptation to network traffic and performance, overcoming the limitations of offline methods that fail to reflect real-world conditions. Our framework thus enables responsive and optimized physical resource block allocation in 5G O-RAN architectures.
To address the aforementioned gaps, we introduce a full-stack O-RAN solution that unifies the OSC near-RT RIC with srsRAN \cite{srsran} for real-time slicing:

\begin{itemize}[left=0pt, noitemsep]
    \item \textbf{End-to-end Integration on OSC RIC and srsRAN:} 
    We develop a near-RT xApp that directly controls srsRAN’s gNB through the E2 interface, using E2AP messages to manage PRB allocations for up to 12 UEs.

    \item \textbf{Fully Online Closed-Loop Training:} 
    Our RL agent receives immediate feedback (downlink throughput, slice QoS) from srsRAN and updates its policy in real time. This ensures a genuine closed-loop approach, where each action on the RAN triggers immediate learning and fine-tuning in the xApp.

    \item \textbf{GNU Radio Channel Emulation:}
    We incorporate Free-Space Path Loss (FSPL), Additive White Gaussian Noise (AWGN), single-tap multipath fading, and Doppler shifts to approximate urban mobility.

    \item \textbf{Practical Insights for Scalability:}
    We highlight operational constraints such as ZeroMQ saturation (limiting concurrency), partial uplink slicing, and attach-sequencing workarounds. These insights inform future O-RAN development for larger-scale deployments.
\end{itemize}

This paper presents an \emph{online}, RL-driven resource allocation strategy for managing multiple slices. 

The system model and E2-based control mechanisms are introduced in Section~\ref{sec:system}, followed by a detailed discussion of the RL methodology and its constraints in Section~\ref{sec:method}. Section~\ref{sec:scenarios} explores the simulation scenarios and channel conditions used to evaluate the framework. The performance of the system under varying mobility and traffic conditions is analyzed in Section~\ref{sec:results}. Finally, Section~\ref{sec:conclusion} summarizes the findings and outlines potential directions for future research in O-RAN.

\ifthenelse{\boolean{longversion}}{
}{}
\section{System Model}
\label{sec:system}

\subsection{Architecture Overview}
\begin{figure}[!t]
    \centering
\includegraphics[width=1\columnwidth, ]{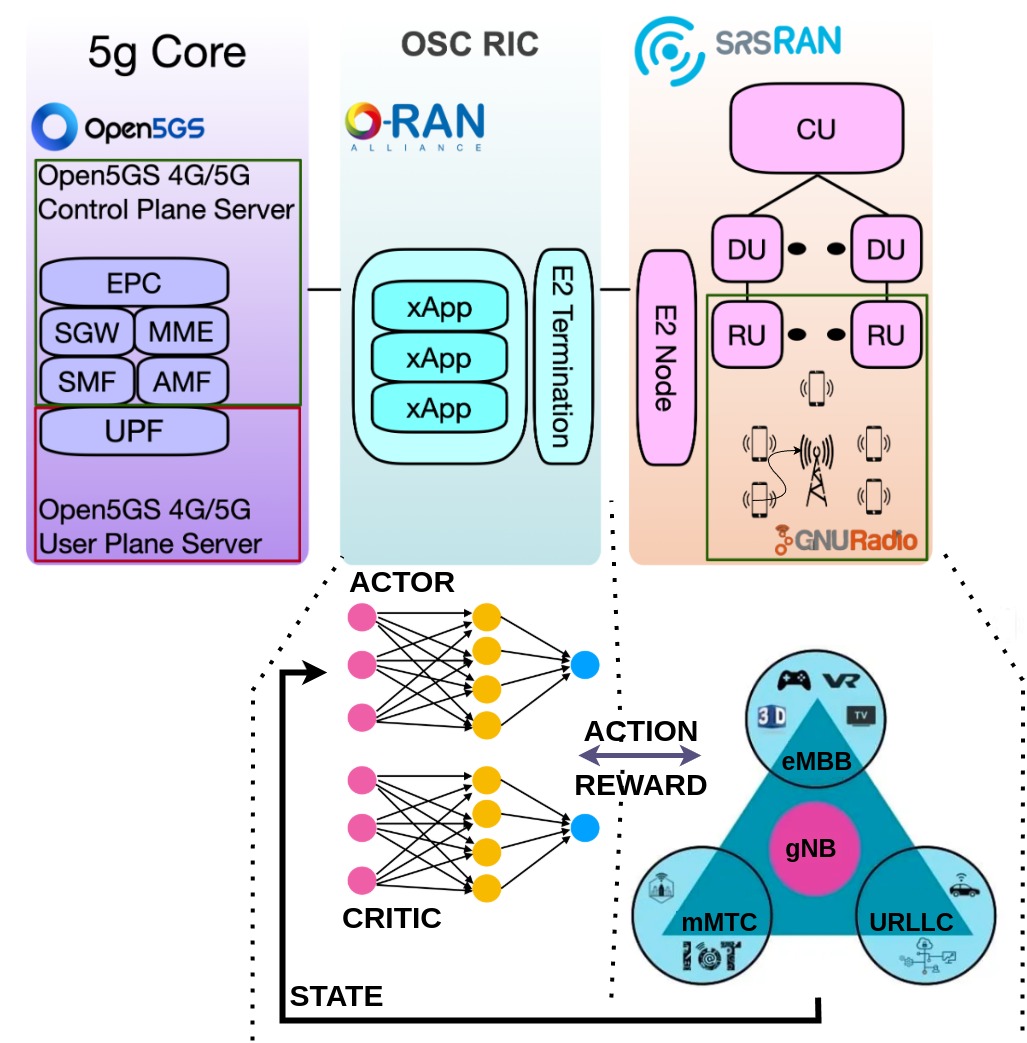}
    \caption{
    Overall system architecture. The near-RT RIC controls the srsRAN gNB via the E2 interface, while GNU Radio injects channel impairments. 
    }
    \label{fig:arch}
\end{figure}

In this work, we integrate the O-RAN Software Community RIC (OSC RIC) with srsRAN’s 5G stack to form a complete end-to-end testbed capable of near-real-time radio resource control. Figure~\ref{fig:arch} illustrates the overall architecture, highlighting the following key components:

\begin{itemize}[left=0pt, noitemsep]
    \item \textbf{Open5GS Core:} Provides both \emph{4G (EPC)} and \emph{5G (5GC)} in a single software suite, enabling seamless UE attachment and data exchange with the RAN.
    \item \textbf{Near-RT RIC:} Hosts \emph{xApps} responsible for coordinating slicing decisions and sending resource-allocation commands via the E2 interface.
    \item \textbf{srsRAN gNB:} Emulates the \emph{Centralized Unit (CU)} and \emph{Distributed Unit (DU)} functionality; an \emph{E2 Service Node (E2SN)} bridges the OSC RIC’s E2Term and the gNB’s internal APIs.
    \item \textbf{E2 Interface:} Enables real-time orchestration of downlink PRB quotas, which can be assigned to multiple slices (URLLC, eMBB, mMTC).
    \item \textbf{GNU Radio Blocks (Channel Model):} Inserted between \emph{srsUE} and \emph{srsGNB} via ZeroMQ streams to emulate path loss, fading, noise, and Doppler effects.
\end{itemize}

In our testbed, \textbf{Open5GS Core} provides the following core network components as Dockerized services using a shared configuration:
\begin{itemize}[left=0pt, noitemsep]
    \item \textbf{EPC (Evolved Packet Core):} Allows 4G operation, with the \emph{MME (Mobility Management Entity)} handling control-plane signaling and the \emph{SGW (Serving Gateway)} routing user-plane data.
    \item \textbf{EPC 5GC (5G Core):} Enables 5G standalone functionality, where the \emph{AMF (Access and Mobility Management Function)} and \emph{SMF (Session Management Function)} manage UE registration and session establishment, and the \emph{UPF (User Plane Function)} forwards user traffic to external networks.
\end{itemize}

A dedicated subscriber database 
contains each UE’s \emph{IMSI}, \emph{encryption keys}, \emph{operator code}, \emph{QoS parameters}, and IP assignment details—referenced by the MME or AMF to authenticate users on network attach. Although our overall testbed supports both \emph{4G} and \emph{5G}, it is specifically deployed in \textbf{5G SA mode} to align with the \emph{3GPP 7.2 architecture} for O-RAN. In a 4G EPC configuration, the MME collaborates with the SGW and PGW (or UPF in a hybrid mode); meanwhile, for 5G SA, the AMF and SMF manage slicing parameters and session establishment, while the UPF routes user-plane traffic externally. \textbf{Network slicing} is enabled by assigning specific \emph{Slice/Service Types (SST)} and \emph{Slice Differentiators (SD)} in the AMF configuration.

\subsection{Channel Assumptions and Modeling}
To ensure realistic network conditions, we incorporate a simplified channel model reflecting an \emph{urban mobility scenario}, accounting for various physical effects:

\begin{enumerate}[left=0pt, noitemsep]
    \item \textbf{Free-Space Path Loss (FSPL):} Signal attenuation is modeled based on the standard free-space path loss formula, which depends on the transmitter--receiver distance, the carrier frequency, and the speed of light. This captures the general propagation loss over distance, typical in open environments.
    
    \item \textbf{Single-Tap Fading:} We introduce a primary path with a complex gain of \([0.85 + 0.25j]\), representing moderate multipath conditions typical of low-rise urban environments. This fading factor is dimensionless, emphasizing the influence of reflections and scattering on signal integrity.
    
    \item \textbf{Additive White Gaussian Noise (AWGN):} Background noise is characterized by a power spectral density proportional to the system temperature, bandwidth, and Boltzmann constant. In practice, the noise voltage is capped below 0.01 in GNU Radio to preserve the stability of the srsRAN implementation.
    
    \item \textbf{Doppler Effects:} For mobile user equipment (UE), such as those supporting Ultra-Reliable Low-Latency Communications (URLLC) at moderate speeds (e.g., 40 km/h), the Doppler shift is computed based on the UE's velocity relative to the carrier frequency. This introduces time-varying phase offsets, presenting challenges for synchronization in dynamic urban scenarios.
\end{enumerate}

The channel modeling is implemented using GNU Radio, which communicates with srsUE and srsGNB via ZeroMQ streams for seamless data exchange. 
In Section~\ref{sec:method}, we elaborate on how these channel characteristics feed into our RL algorithm’s observation space.


\section{Methodology}
\label{sec:method}

\subsection{Reinforcement Learning Framework for RAN Slicing Optimization}

We employ Proximal Policy Optimization (PPO) which is an actor-critic framework to address the decision-making challenges of RAN slicing. Here, \(s_t\) represents the state at timestep \(t\), which encapsulates the current UE requirements and status. The actor network outputs action vectors \(\mathbf{a}_t\), representing PRB allocations for each slice at timestep \(t\), balancing resource provisioning and over-allocation. The critic network estimates the value function \(V^\pi(\mathbf{s}_t)\), where \(\pi\) denotes the policy, a mapping from states \(s_t\) to actions \(a_t\), which the agent learns to optimize the long-term reward.

\subsubsection{State and Action Spaces}
The state vector comprises
of UE type, Bit-rate or Buffer size (occupancy) and pathloss, relying primarily on E2 metrics provided by srsRAN. The action space defines PRB allocations at the slice level (eMBB, URLLC, mMTC), rather than per UE. PRBs are first distributed evenly among UEs within each slice, with any remaining PRBs allocated to UEs experiencing higher path loss. The total PRB usage remains constrained by the subframe limit, ensuring compliance with srsRAN’s channel bandwidth.

\subsubsection{Slice-Specific Reward Functions}
\label{subsec:rewards}

We design a reward function $R_t$ that combines the individual rewards $r_{\text{URLLC}, t}$, $r_{\text{eMBB}, t}$, and $r_{\text{mMTC}, t}$ at time $t$. Each slice $s \in \{\text{URLLC, eMBB, mMTC}\}$ has a distinct QoS priority and thus a specialized sub-reward. The final reward is a weighted sum:
\begin{align}
R_t \;=\; \alpha_{\text{URLLC}} \,r_{\text{URLLC},t}
\;+\;\alpha_{\text{eMBB}}\,r_{\text{eMBB},t}
\;+\;\alpha_{\text{mMTC}}\,r_{\text{mMTC},t},
\end{align}
where $\alpha_{\text{URLLC}}, \alpha_{\text{eMBB}}, \alpha_{\text{mMTC}}$ are slice importance weights. Below, we define each sub-reward:

\paragraph{URLLC Reward ($r_{\text{URLLC}}$).}
URLLC focuses on ultra-low latency and reliability. We approximate latency by monitoring buffer occupancy or per-packet delay. We define a negative penalty if the slice’s average delay exceeds $t_{\text{target}}$:
\begin{align}
r_{\text{URLLC}, t} \;=\;
\max\!\Bigl(-1,\,\min\!\Bigl(0,\,
\dfrac{t_{\text{target}} - t_{\text{avg}}}{t_{\text{target}}}\Bigr)\Bigr),
\end{align}
where $t_{\text{avg}}$ is the measured or estimated delay (e.g., from buffer timestamps), and $t_{\text{target}}$ is the URLLC delay threshold. Lower delay yields higher (less negative) reward.

\paragraph{eMBB Reward ($r_{\text{eMBB}}$).}
eMBB aims for high data throughput. We track the slice’s average bitrate $b_{\text{avg}}$ over the last measurement window and compare it against a target $b_{\text{target}}$:
\begin{align}
r_{\text{eMBB}, t} \;=\;
\max\!\Bigl(-1,\,\min\!\Bigl(0,\,
\dfrac{b_{\text{avg}} - b_{\text{target}}}{b_{\text{target}}}\Bigr)\Bigr).
\end{align}
If $b_{\text{avg}}$ meets $b_{\text{target}}$, the sub-reward approaches 0. Otherwise, it becomes increasingly negative, with a floor at $-1$.

\paragraph{mMTC Reward ($r_{\text{mMTC}}$).}
For mMTC, we focus on the number of successfully received packets—especially if the devices are mostly downlink or rely on sporadic transmissions. Let $b_{\text{received}}$ be the total bytes (or packets) received and $b_{\text{expected}}$ the desired or generated amount over the window:
\begin{align}
r_{\text{mMTC}, t} \;=\;
\max\!\Bigl(-1,\,\min\!\Bigl(0,\,
\dfrac{b_{\text{received}} - b_{\text{expected}}}{b_{\text{expected}}}\Bigr)\Bigr).
\end{align}
When $b_{\text{received}} \approx b_{\text{expected}}$, the slice gets a near-zero penalty. If $b_{\text{received}}$ is too low, it is penalized more harshly.

\paragraph{Clipping and Weights.}
We constrain each slice sub-reward to $[-1,\,0]$ to avoid exploding gradients and to unify the magnitude of slice-specific penalties. The weighting factors $\alpha_{\text{URLLC}}, \alpha_{\text{eMBB}}, \alpha_{\text{mMTC}}$ allow fine-tuning of slice priority and configurable fairness. For example, if URLLC must absolutely not be violated, one might set $\alpha_{\text{URLLC}} \gg \alpha_{\text{eMBB}}$.

This formulation ensures each slice’s QoS goals (latency, throughput, or packet reception) are explicitly represented within a single scalar reward $R_t$. During training, the RL agent learns how to adjust PRB allocations across slices to maximize $R_t$ over time, striking a balance among slice priorities and the channel constraints.

\subsubsection{Training Procedure}
The training procedure is as follows: the RL agent sets resource allocation while the xApp applies PRB assignments. During traffic simulation, KPI measurements are gathered every 500 ms. After simulation, the xApp averages these KPIs and saves them in a JSON file. The RL agent then computes the reward using the KPIs and stores the state, action, and reward in replay memory to update its policy. Table \ref{tab:task_spec} lists the parameters for each slice.



\begin{table}[h]
    \centering
    \renewcommand{\arraystretch}{1.2}
    \setlength{\tabcolsep}{6pt}
    \begin{tabular}{c|c|c}
        \hline
        \textbf{Service} & \textbf{Parameter} & \textbf{Values} \\
        \hline
        \multirow{4}{*}{URLLC} & Gen. Freq. (Hz) & 2 \\
                               & Gen. Bytes (B) & Min: $10^5$, Max: $3\times10^5$ \\
                               & Latency (ms) & 500 \\
                               & $\alpha_{\text{URLLC}}$ & 1 \\
        \hline
        \multirow{2}{*}{eMBB} & Bit Rate & Min: $2\times10^5$, Max: $4\times10^5$ \\
                              & $\alpha_{\text{eMBB}}$ & 1 \\
        \hline
        \multirow{3}{*}{mMTC} & Gen. Freq. (Hz) & 4 \\
                              & Gen. Bytes (B) & Min: $25\times10^3$, Max: $60\times10^3$ \\
                              & $\alpha_{\text{mMTC}}$ & 1 \\
        \hline
    \end{tabular}
    \caption{Task Specification for PRB Allocation}
    \label{tab:task_spec}
\end{table}

\subsection{Resource Management Constraints: Scalability Limitations and Open-Source Commitment}

While our RL approach is designed to be generic and adaptable to diverse network conditions, 
the limited computational resources on one PC would propose a limitation.
We run our environment via virtualization on a 24-core, 32-thread Intel(R) Core(TM) i9-14900K CPU. In this setup, ZeroMQ, being CPU-bounded and operating as a lightweight TCP/IP-based point-to-point transport layer, manages inter-process communication between srsGNB and each simulated srsUE. Although GNU Radio connects these individual message streams, the use of ZeroMQ in this design would propose a communicational limit. While it supports N-to-N socket instances, increasing the number of simultaneous endpoints 
causes network strain, buffer overflow, and eventual saturation.
Extensive testing shows that an unsaturated channel in this environment supports a total of 28 Mbps across all UEs. However, when more than three UEs transmit simultaneously, the channel saturates, leading to substantial performance degradation and system instability. 
To mitigate this constraint and enable simulations with up to 12 UEs, we employ a batching strategy that groups UEs in sets of three, rotating these groups over time intervals as described in Section~\ref{subsec:scenarios-limits}. It is worth noting that more CPU cores would likely improve performance.

In contrast, 
proprietary solutions like \textit{Amarisoft}\cite{amarisoft} exhibit superior scalability, supporting significantly more UEs without similar bottlenecks, but their closed-source nature restricts reproducible research efforts. Our work underscores the importance of open-source frameworks to foster collaborative O-RAN development. Additionally, as of now, the current stable version of srsRAN supports only down-link slicing
We address these challenges by leveraging the O-RAN E2 interface in a closed-loop framework with our resource allocation xApp, enabling dynamic PRB allocation and near real-time KPI monitoring directly over E2. Although throughput measurements are sampled every 500 ms,
this setup provides timely feedback for the RL agent to update its policy, adapt resource allocations, and evaluate the impact of state changes. 
the RL agent relies on higher-level indicators like throughput and slice occupancy, restricting its ability to optimize network performance effectively. 
From the KPI perspective, the RL agent relies on throughput readings.
Despite the constraints, our methodology demonstrates the feasibility of real-time RL-based slicing with srsRAN. 

\section{Simulation Scenarios}
\label{sec:scenarios}

\subsection{ Traffic and Deployment Configuration}

We design our simulation scenarios around three primary slices—{URLLC}, {eMBB}, and {mMTC}—each reflecting distinct downlink-centric 5G services with corresponding QoS demands. The {URLLC} slice (\(4\) UEs) is modeled as autonomous driving systems requiring ultra-reliable, low-latency transmissions at \(40\,\mathrm{km/h}\), emulated by large, non-frequent downlink packets. The {eMBB} slice (\(4\) UEs) comprises two mobile UEs representing smartphone users streaming high-bandwidth content on the move, and two stationary UEs simulating fixed wireless access subscribers consuming large data flows (e.g., 4K video). Finally, the {mMTC} slice (\(4\) UEs) mimics low-throughput IoT endpoints such as digital billboards, where small, high frequency downlink updates reflect srsRAN’s limited uplink support.

In total, the deployment accommodates up to \(12\) UEs, though only three transmit actively at any given time to avoid ZeroMQ saturation. 
These active UEs rotate in groups, maintaining stable throughput while capturing dynamic traffic patterns. 
A realistic channel model—including single-tap path loss, AWGN, Doppler effects, and random distances (\(d \in [0.5\text{\,km}, 2\text{\,km}]\))—emulates a mobile urban environment. This setup provides an efficient testbed for evaluating RL-based resource allocation strategies under diverse, downlink-centric QoS requirements.


\subsection{Overcoming Scalability Constraints in Open Source Deployments}
\label{subsec:scenarios-limits}

 \ifthenelse{\boolean{longversion}}{

%
}
Since \texttt{srsRAN}, originally designed for hardware-SDR use, one may face few issues while scaling up the number of simulated UEs in a purely software-based digital twin.
We aim to keep our solution accessible for the research community and thus outline the following key challenges in replicating our efforts:

\begin{enumerate}[label=\textbf{(\arabic*)}, leftmargin=1.5em]
  \item \textbf{PRACH Attach Delays} \\
  Simultaneous attachments overwhelm \texttt{ZeroMQ} buffers, causing crashes. To mitigate this, each \texttt{srsUE} adds a random PRACH offset, staggering attach attempts.

  \item \textbf{Downlink-Only Slicing} \\
  The current stable version of \texttt{srsUE} as of writing this paper only supports downlink slicing. While adequate for downlink use cases (e.g., video streaming), this excludes the stringent uplink demands of URLLC and mMTC.

  \item \textbf{Batching Transmitting UEs} \\
  Activating more than three \texttt{srsUE} instances concurrently saturates ZeroMQ and results in a channel throughput degradation. 
  To address this, only three UEs transmit simultaneously.
\end{enumerate}

\subsection{Evaluation Metrics and Logging}
\label{subsec:scenarios-eval}
Performance is evaluated using per-UE E2 throughput at a \(500\,\mathrm{ms}\) sampling rate.
Training the agent consists of two phase.
The first phase is the pre-training phase which utilizes the channel throughput we accomplished during our experiments with different number if PRBs in the simulator, rather than simulating the traffic. We use this pre-training phase so that the RL agent has a good starting weights for online learning inside the simulator.
This model is then integrated into the simulator for the second training phase, enabling the agent to learn performance implications under realistic constraints such as FSPL, AWGN, fading, and frequency drift. 

Algorithm \ref{PPO_PRB_Allocation} outlines the procedural steps for PPO-based PRB allocation. The training process begins with the initialization of the PPO agent, where policy and value function parameters are defined. During each training iteration, new tasks are generated for all UEs, and the corresponding state vector is constructed and provided to the PPO agent. Based on the observed state, the PPO agent selects an action that determines the PRB allocation for each slice. The xApp subsequently applies the allocation through RAN Controller (RC) messages and UEs start generating traffic. The KPIs are continuously monitored and logged to evaluate the system's performance. The observed KPIs are used to compute a reward, which is stored along with the state-action pair in the experience buffer. When an update step is reached, the algorithm computes the advantage estimate using Generalized Advantage Estimation (GAE), optimizes the policy via clipped surrogate loss, and updates the value function through regression on discounted returns. The training process continues iteratively until the convergence criteria are met.

\begin{algorithm}
\caption{PPO-based PRB Allocation}
\label{PPO_PRB_Allocation}
\begin{algorithmic}[1]  

    \State \textbf{Initialize:} PPO agent with policy parameters $\theta$ and value function parameters $\phi$

    \While{Training is not complete}
        \State Generate new tasks for all UEs
        \State Construct the state vector $\mathbf{s}_t$
        \State Provide state $\mathbf{s}_t$ to the PPO agent
        \State PPO agent selects action $a_t \sim \pi_{\theta}(a_t | \mathbf{s}_t)$
        \State Allocate PRBs to UEs based on action $a_t$
        \State xApp applies PRB allocation via RC messages
        \State UEs generate traffic
        \State xApp monitors KPIs and logs data in a JSON file
        \State Compute reward $r_t$ based on observed KPIs
        \State Store transition $(\mathbf{s}_t, a_t, r_t)$ in experience buffer
        
        \If {Update step is reached}
            \State Compute advantage estimate $\hat{A}_t$ using Generalized Advantage Estimation (GAE)
            \State Optimize policy $\pi_{\theta}$ using the clipped surrogate loss function
            \State Update value function $V_{\phi}$ 
            \State Perform gradient ascent on $\theta$ and $\phi$
        \EndIf
    \EndWhile

\end{algorithmic}
\end{algorithm}

One of the key advantages of using online training for PPO in this context is its ability to dynamically adapt to changing network conditions. Unlike offline training, which relies on precollected data, online training enables the agent to continuously refine its policy based on the interaction it has with the environment. This adaptability is crucial in O-RAN environments, where traffic patterns and resource demands fluctuate dynamically. However, this approach comes with a significant computational cost. 

\section{Quality of Service Analysis}
\label{sec:results}

In this section, we evaluate the performance of our RL-based slicing framework under the simulation scenarios described in Section~\ref{sec:scenarios}. Our analysis focuses on the overall QoS compliance and the stability of the proposed framework.

We compare the performance of our PPO-based resource allocation approach, detailed in Section~\ref{sec:method}, against DQN and four baseline methods. To assess resource allocation efficiency, we utilize the cumulative distribution function (CDF) to analyze key performance metrics. Specifically, we examine the latency distribution of URLLC traffic to determine compliance with stringent delay requirements. Additionally, for eMBB services, we evaluate the difference between the achieved downlink bitrate and the requested bitrate. For mMTC, we assess the discrepancy between the number of received bytes and the actual transmitted bytes.

Furthermore, we analyze the impact of network congestion and resource contention on service quality, highlighting the ability of each strategy to meet dynamic service demands. Finally, we evaluate overall QoS compliance by measuring adherence to throughput and latency targets across network slices. This includes examining the trade-offs between latency-sensitive URLLC traffic and the bandwidth-intensive requirements of eMBB and mMTC services, providing insights into the effectiveness of resource allocation under varying network conditions.

For benchmarking, we compare the RL-based methods with four baseline resource allocation strategies:

\begin{itemize}[left=0pt, noitemsep]
    \item \textbf{Equal Allocation Baseline:} This method distributes the total number of PRBs equally among all active users using integer division. This approach provides a straightforward and fair division of resources but does not adapt dynamically to traffic demands or channel conditions.
    
    \item \textbf{Proportional Allocation Baseline:} This method assigns PRBs in proportion to user demand, which is determined differently based on the traffic type. For URLLC and mMTC users, demand is computed as the product of generation frequency and packet size, assuming a fixed generation period. PRBs are allocated according to each user's share of the total demand, with any leftover PRBs assigned to users with the largest fractional remainder. This approach ensures a more adaptive distribution of resources based on instantaneous demand.
    
    \item \textbf{Pre-allocated Proportional Baseline:} This method ensures that each user receives at least one PRB before distributing the remaining resources based on proportional demand. This method prevents extreme resource starvation by guaranteeing a minimum allocation while still adapting to demand variations.
    
    \item \textbf{3GPP-Based Proportional Fair Frequency-Domain Packet Scheduling (3GPP-PF) \cite{5062197}:} This method follows the proportional fair scheduling principle standardized in 3GPP, balancing user fairness and spectral efficiency by prioritizing users based on their channel conditions and past resource allocations. The proportional fair scheduler dynamically adjusts allocations to mitigate resource starvation and enhance overall spectral efficiency.
\end{itemize}

\begin{figure*}[!t]
    \centering
    \begin{subfigure}[b]{0.31\textwidth}
        \centering
        \includegraphics[width=\linewidth]{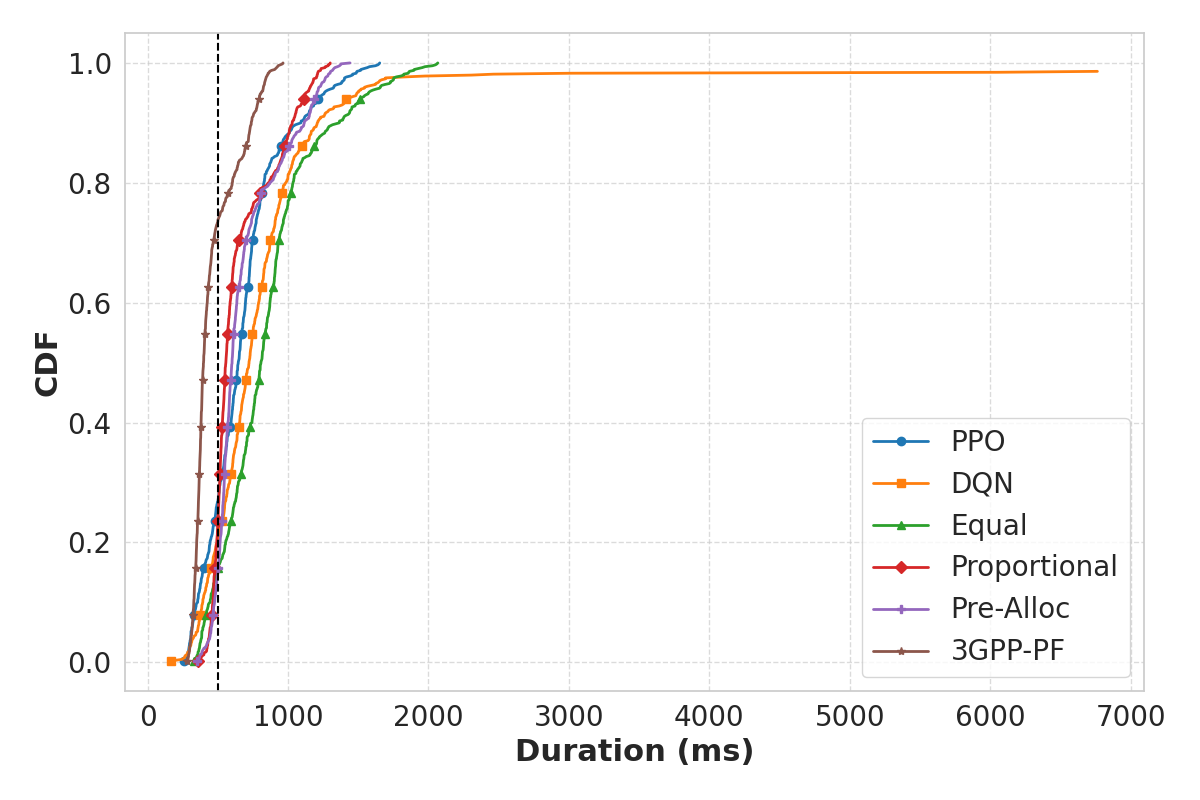}
        \caption{CDF of URLLC Latency.}
        \label{fig:urlcc_cdf}
    \end{subfigure}
    \hfill 
    \begin{subfigure}[b]{0.31\textwidth}
        \centering
        \includegraphics[width=\linewidth]{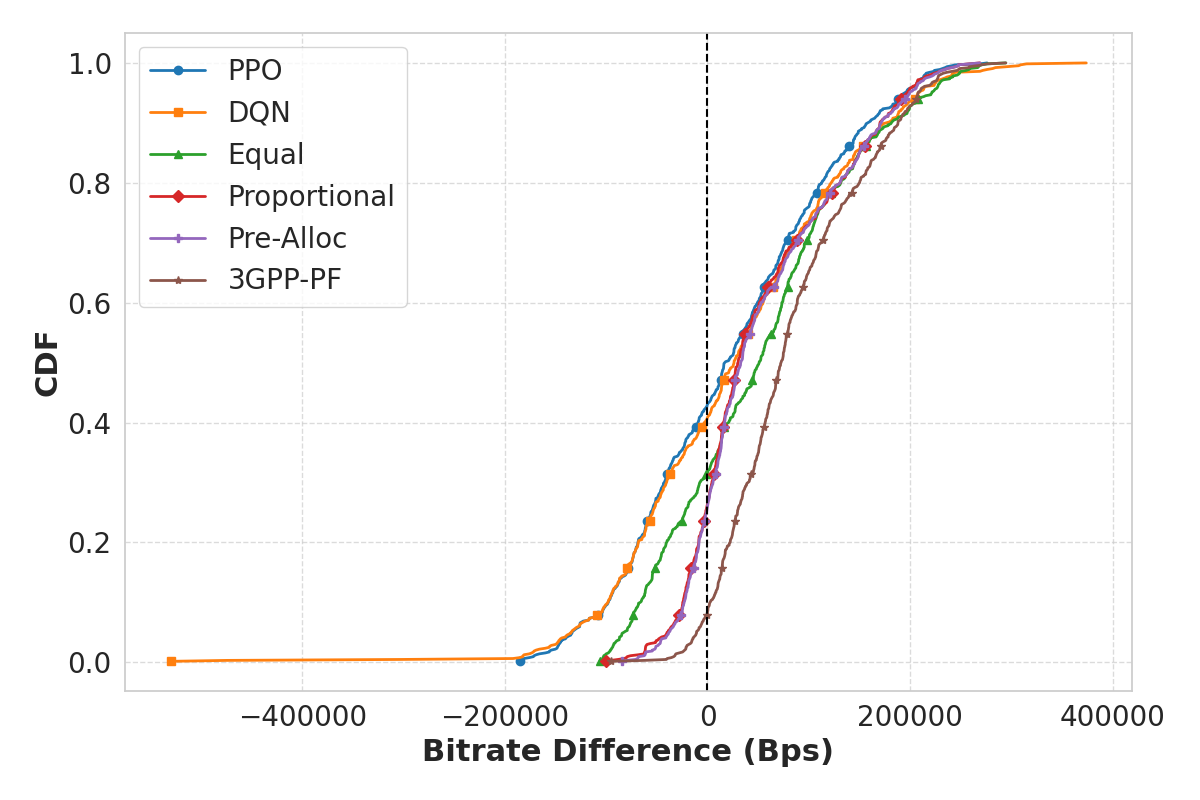}
        \caption{CDF of eMBB $\Delta$ Bitrate.}
        \label{fig:embb_cdf}
    \end{subfigure}
    \hfill
    \begin{subfigure}[b]{0.31\textwidth}
        \centering
        \includegraphics[width=\linewidth]{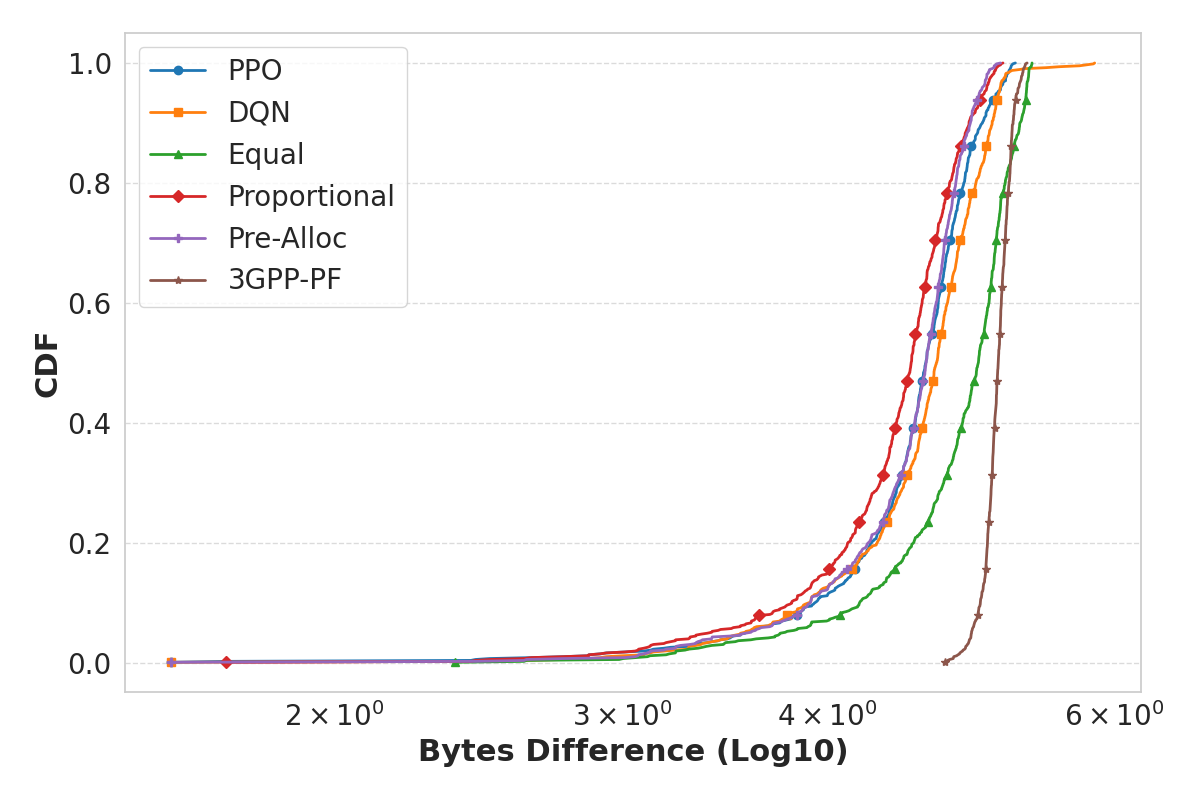}
        \caption{CDF of mMTC $\Delta$ Payload (Log Scale).}
        \label{fig:mmtc_cdf}
    \end{subfigure}
    \caption{Comparison of network slice CDFs.}
    \label{fig:three_cdfs}
\end{figure*}

The evaluation of the CDF plots, shown in Figures \ref{fig:urlcc_cdf}, 
\ref{fig:embb_cdf} and ~\ref{fig:mmtc_cdf} highlights the comparative performance of the RL-based approaches (PPO and DQN) against the baseline resource allocation methods.
In Figure~\ref{fig:mmtc_cdf}, which depicts the bitrate distribution for mMTC slices, PPO demonstrates superior performance by achieving higher throughput levels for a larger proportion of users compared to DQN and the baselines. Similarly, in Figure~\ref{fig:embb_cdf}, which illustrates the throughput distribution for eMBB slices, PPO consistently outperforms other methods, indicating its ability to allocate resources adaptively to meet the higher bandwidth demands of eMBB traffic.
For URLLC traffic, as shown in Figure~\ref{fig:urlcc_cdf}, PPO achieves lower latency for a greater percentage of packets, ensuring compliance with stringent delay requirements. Specifically, the latency CDF indicates that PPO prioritizes low-latency traffic more effectively by dynamically allocating PRBs to meet the strict timing needs of URLLC packets, even under fluctuating network conditions and resource contention with other traffic types. This is particularly evident in the steepness of the latency curve for PPO, which reflects a concentrated distribution of packets with delays well below the critical thresholds. Compared to DQN and the baselines, PPO exhibits a more robust adaptation to channel variability, maintaining consistent performance across different scenarios. While the pre-allocated proportional baseline provides a reasonable minimum latency for many packets due to its fairness-driven approach, its inability to dynamically adjust to traffic demands results in a heavier tail, where a non-negligible percentage of packets experience higher delays. The 3GPP-PF approach, while effective in maintaining fairness, is limited by its reliance on past allocations, leading to occasional inefficiencies in handling dynamic traffic loads. Overall, PPO's latency optimization demonstrates its capability to fulfill the ultra-reliable and low-latency requirements critical for URLLC applications, ensuring that stringent QoS objectives are met.




\section{Conclusion}
\label{sec:conclusion}
In this paper, we presented a framework integrating the OSC near-RT RIC with srsRAN for real-time slicing and resource management in O-RAN. Using PPO in an RL-based xApp, we demonstrated adaptive resource allocation for URLLC, eMBB, and mMTC slices under realistic channel conditions modeled with GNU Radio. Our results show improved QoS compliance, including enhanced throughput, reduced latency, and stable resource distribution.

Despite these achievements, several challenges remain. One key limitation is the high computational demand of PPO, particularly in online training scenarios where policies must dynamically adapt to real-time network variations. This contrasts with offline training methods that rely on pre-collected datasets, reducing computational overhead but limiting adaptability.
Additionally the constraints imposed by not using SDRs particularly in handling increased numbers of UEs due to ZeroMQ.
Moreover, our work underscores the importance of balancing URLLC's stringent latency requirements with eMBB's high throughput demands. The ability of PPO to dynamically reallocate resources based on evolving network conditions proves advantageous, but additional refinements in state-space representation and reward design could enhance performance further. Future research should explore hybrid approaches combining online and offline RL training to mitigate computational burdens while preserving adaptability. Additionally, integrating multi-agent RL techniques could improve decision-making scalability in dense O-RAN deployments.

Overall, this work establishes a foundation for real-time RL-based slicing in O-RAN while identifying areas for future enhancement. Addressing these limitations will be crucial in achieving robust, scalable, and intelligent resource allocation strategies for next-generation wireless networks.

\ifthenelse{\boolean{longversion}}{
}{}


\end{document}